\documentclass[sigconf, screen]{acmart}

\setcopyright{rightsretained}
\acmDOI{10.1145/3650212.3652141}
\acmYear{2024}
\copyrightyear{2024}
\acmSubmissionID{issta24main-p1180-p}
\acmISBN{979-8-4007-0612-7/24/09}
\acmConference[ISSTA '24]{Proceedings of the 33rd ACM SIGSOFT International Symposium on Software Testing and Analysis}{September 16--20, 2024}{Vienna, Austria}
\acmBooktitle{Proceedings of the 33rd ACM SIGSOFT International Symposium on Software Testing and Analysis (ISSTA '24), September 16--20, 2024, Vienna, Austria}
\received{16-DEC-2023}
\received[accepted]{2024-03-02}

\AtBeginDocument{%
  \providecommand\BibTeX{{%
    \normalfont B\kern-0.5em{\scshape i\kern-0.25em b}\kern-0.8em\TeX}}}

\newcommand{\sys}{WasmRev} 
\newcommand{\squishlist}{
   \begin{list}{$\bullet$}{%
        \setlength{\itemsep}{0pt}%
        \setlength{\parsep}{0pt}%
        \setlength{\topsep}{0pt}%
        \setlength{\partopsep}{0pt}%
        \setlength{\listparindent}{-2pt}%
        \setlength{\itemindent}{-5pt}%
        \setlength{\leftmargin}{1.2em}%
        \setlength{\labelwidth}{0em}%
        \setlength{\labelsep}{0.5em}%
    }
}
\newcommand{\squishend}{
    \end{list}  }
\usepackage{microtype}
\usepackage[skip=5pt]{caption}
\usepackage{xcolor}
\usepackage{comment}
\usepackage{xspace}
\usepackage[utf8]{inputenc}
\usepackage[T1]{fontenc}
\usepackage{tabu}
\usepackage{tabularx}
\usepackage{tablefootnote}
\usepackage{threeparttable}
\usepackage{algorithm}
\usepackage{algpseudocode}
\algtext*{EndFor}
\algtext*{EndIf}
\usepackage[T1]{fontenc}
\usepackage{contour}
\usepackage{amsmath}
\usepackage{booktabs}
\usepackage{dsfont}
\usepackage{mfirstuc}
\usepackage{multirow}
\usepackage{listings}
\usepackage{float}
\usepackage{subcaption}
\usepackage{tikz}
\usepackage{boldline}
\usepackage{balance}
\usepackage{lipsum} 
\usepackage[skins]{tcolorbox}%
\tcbuselibrary{breakable}%
\tcbuselibrary{hooks}

\begin{document}

\title{Multi-modal Learning for WebAssembly Reverse Engineering}

\author{Hanxian Huang}
\orcid{0000-0001-6338-3289}
\affiliation{%
  \institution{University of California at San Diego}
  \city{San Diego}
  \country{USA}
}
\email{hah008@ucsd.edu}

\author{Jishen Zhao}
\orcid{0000-0002-8766-0946}
\affiliation{%
  \institution{University of California at San Diego}
  \city{San Diego}
  \country{USA}
}
\email{jzhao@ucsd.edu}

\begin{abstract}
The increasing adoption of WebAssembly (Wasm) for performance-critical and security-sensitive tasks drives the demand for WebAssembly program comprehension and reverse engineering. 
Recent studies have introduced machine learning (ML)-based WebAssembly reverse engineering tools. Yet, the generalization of task-specific ML solutions remains challenging, because their effectiveness hinges on the availability of an ample supply of high-quality task-specific labeled data. Moreover, previous works trained models only with features extracted from WebAssembly, overlooking the high-level semantics present in the corresponding source code and its documentation. Acknowledging the abundance of available source code with documentation, which can be compiled into WebAssembly, we propose to learn representations of them concurrently and harness their mutual relationships for effective WebAssembly reverse engineering. 

In this paper, we present \sys{}, the first multi-modal pre-trained language model for WebAssembly reverse engineering. \sys{} is pre-trained using self-supervised learning on a large-scale multi-modal corpus encompassing source code, code documentation and the compiled WebAssembly, without requiring labeled data. \sys{} incorporates three tailored multi-modal pre-training tasks to capture various characteristics of WebAssembly and cross-modal relationships. \sys{} is only trained once to produce general-purpose representations that can broadly support WebAssembly reverse engineering tasks through few-shot fine-tuning with much less labeled data, improving data efficiency. We fine-tune \sys{} onto three important reverse engineering tasks: type recovery, function purpose identification and WebAssembly summarization. Our results show that \sys{} pre-trained on the corpus of multi-modal samples establishes a robust foundation for these tasks, achieving high task accuracy and outperforming the state-of-the-art ML methods for WebAssembly reverse engineering.

\end{abstract}


  
\maketitle

\section{Introduction}




WebAssembly (Wasm) is a low-level, portable, bytecode format compiled from high-level languages, such as C, C++, and Rust, delivering near-native performance when executed on the web~\cite{haas2017bringing,WebAssembly_website}. It is a promising technology to enhance performance of various applications traditionally developed in JavaScript, e.g., cryptography~\cite{attrapadung2018efficient,ctwasm} 
and machine learning~\cite{TF_js,wasm_onnx}. 
Initially conceived for execution in web browsers, WebAssembly is now expanding its reach to encompass various application domains, such as serverless computing~\cite{severless,gadepalli2019challenges_severless}, edge computing~\cite{gadepalli2020sledge}, and IoT~\cite{wen2020wasmachineIOT,makitalo2021webassemblyIOT}. 

As WebAssembly gains popularity across diverse applications, an increasing demand emerges for understanding and reverse engineering WebAssembly code. Many WebAssembly modules -- including potentially malicious ones -- are distributed through third-party services, rendering the source code unavailable on the client-side~\cite{musch2019new}. This requires users to understand the WebAssembly modules and audit it to prevent potential attacks or malicious code~\cite{lehmann2020everything,wang2018seismic,romano2020minerray}. While WebAssembly has a readable text format, manual interpretation and comprehension remain challenging and error-prone. Different from register-based native binaries (e.g., x86 or ARM), WebAssembly adopts a stack machine design, necessitating the tracing of stack behavior to comprehend the code or compute a specific variable. Moreover, WebAssembly only employs four numeric data types, \textsf{i32}, \textsf{i64}, \textsf{f32,} and \textsf{f64}, which conceals the data type information and further increases the complexity of comprehending WebAssembly.

Several prior studies explored WebAssembly understanding and reverse engineering with conventional analysis approaches~\cite{brito2022wasmati,jeong2018watt,fu2018taintassembly,szanto2018taint} or machine learning (ML)-based methods~\cite{lehmann2022finding, romano2023automated}. 
Traditional approaches analyze WebAssembly through precise, logical reasoning, often incorporating heuristics to ensure the practical utility of the tools~\cite{brito2022wasmati,jeong2018watt,fu2018taintassembly,szanto2018taint}. However, crafting effective heuristics is difficult, especially in cases where an accurate analysis result depends on uncertain information, such as natural language (NL) content in code documentation and NL code search tasks~\cite{feng2020codebert,lu2021codexglue,husain2019codesearchnet}. This uncertainty is not conductive to logic-based reasoning~\cite{NeuralSE}.

\begin{figure}[t!]
  \centering
  \includegraphics[width=0.62\linewidth]{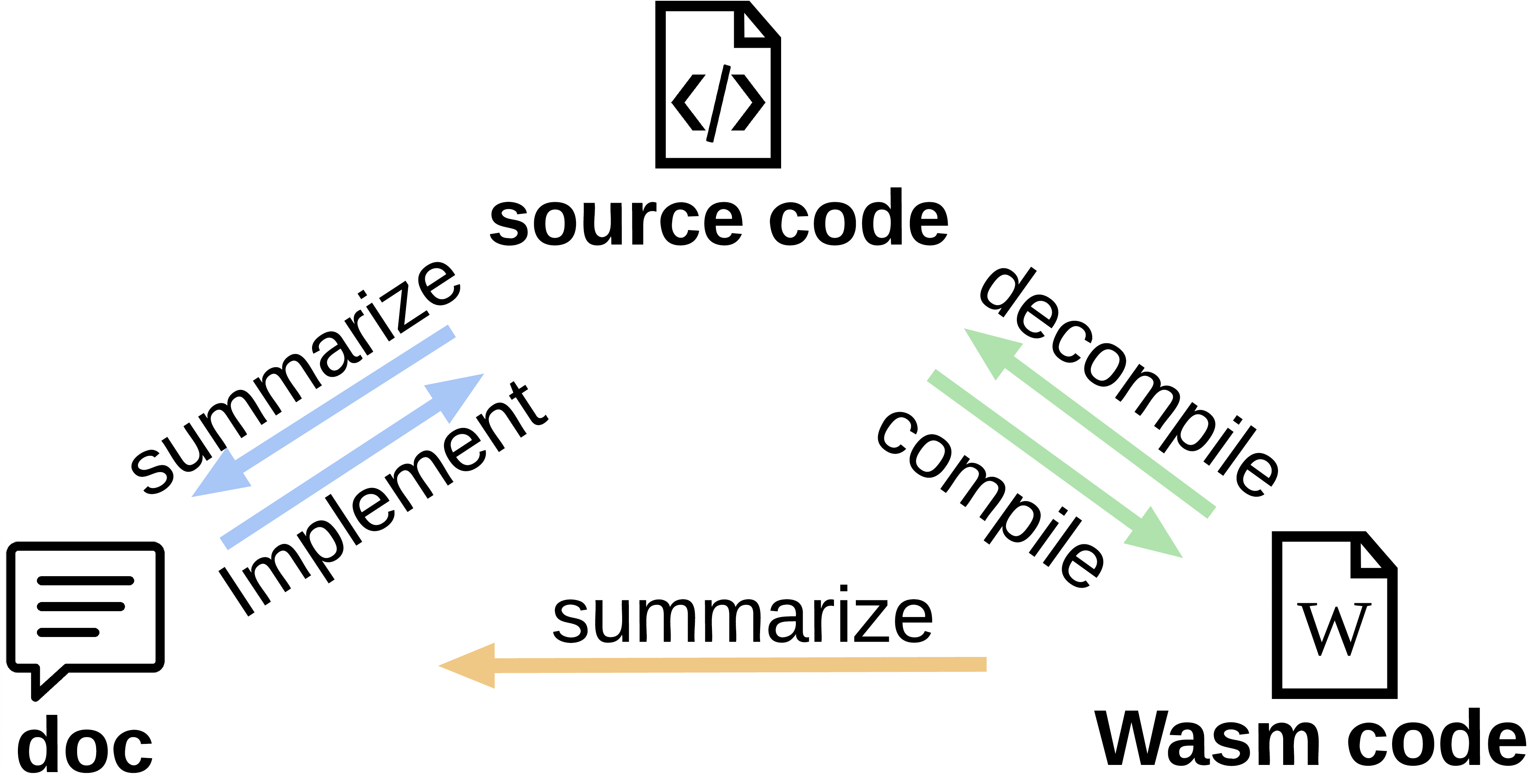}
  \caption{The relationship of engineering and reverse engineering among source code, code documentation, and WebAssembly code.}
  \vspace{-10pt}
  \label{fig:relationship}
\end{figure}

Instead, recent 
ML-based solutions
learn features from extensive labeled code datasets and manipulate fuzzy information in code without relying on manually encoded precise rules. SnowWhite~\cite{lehmann2022finding} trains a sequence model based on WebAssembly code 
to recover precise types. 
WASPur~\cite{romano2023automated} is trained on control flow graph features from WebAssembly code to identify function purpose. However, one key caveat is that these frameworks are designed for specific tasks, while overlooking the generic WebAssembly features shared across various WebAssembly analysis tasks. In addition, both frameworks require large scale task-specific labeled datasets, which are laborsome to collect for many analysis tasks. For instance, function purpose identification requires manual inspection of WebAssembly code, checking high-level semantics like function names to label the correct purpose of a function~\cite{romano2023automated}. Hence, generalizing and applying these task-specific models to a broader range of WebAssembly tasks poses a significant challenge.

Despite substantial efforts in training ML models for WebAssembly~\cite{lehmann2022finding,romano2023automated}, the relationships among high-level source code, code documentation, and WebAssembly code remain crucial yet underexplored in the realm of WebAssembly reverse engineering. As depicted in Figure~\ref{fig:relationship}, a program is often composed of both code snippets (in a high-level programming language \textbf{(PL)}) and corresponding documentation (in natural language \textbf{(NL)}), while the code snippets can be compiled into WebAssembly (\textbf{Wasm}). In this paper, we define source code, code documentation, and WebAssembly code as multiple modalities of code -- \textbf{(NL, PL, Wasm)}. The source code (PL) serves as a human-readable representation of the intended logic of a program in a high-level programming language; the code documentation (NL) supplements this by offering insights into the purpose of functions in natural language. Studying both offers contextual information that aids in interpreting the purpose and behavior of the code, providing valuable knowledge for reverse engineers seeking a comprehensive understanding and interpretation of 
WebAssembly code. 
As such, 
concurrently learning representations of these semantically equivalent modalities, along with understanding their relationships, will yield complementary information critical for comprehending WebAssembly.

To address the aforementioned challenges and limitations, we propose \sys{}, the first multi-modal pre-trained language model for WebAssembly reverse engineering. Figure~\ref{fig:workflow} shows an example of \sys{} workflow. 
Different from previous task-specific models~\cite{lehmann2022finding,romano2023automated}, \sys{} effectively learns a generic representation among WebAssembly code, source code, and code documentation, and efficiently transfers it to perform various WebAssembly reverse engineering tasks. 
To this end, we tailor three pre-training tasks enabling \sys{} to effectively learn inter-modal and intra-modal relationships, and close the gaps across various modalities of representations (Section~\ref{sec:pre-training}). The pre-training is self-supervised, without requiring a large scale labeled dataset, effectively solving the inherent scarcity of labeled dataset issues. The pre-training is done once, while the pre-trained \sys{} can be shared and adapted to various WebAssembly tasks with light-weight fine-tuning. Compared with supervised learning on specific tasks, such a process is more efficient and requires much less task-specific labeled data (Section~\ref{sec:fine-tuning}). \sys{} offers efficient and accurate solutions for various WebAssembly reverse engineering tasks, furnishing programmers with insightful high-level abstractions to enhance an accessible comprehension of WebAssembly and facilitate further code inspection. 
In summary, this paper contributes the following:

\begin{figure}[t!]
  \centering
  \includegraphics[width=\linewidth]{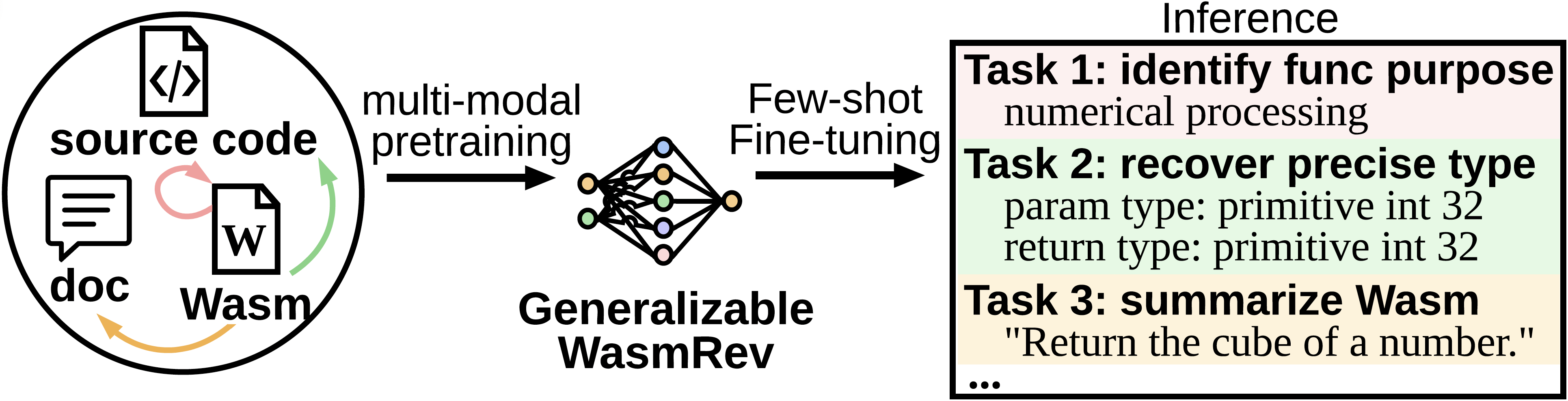}
  \caption{\sys{} overview.}
  \vspace{-8pt}
  \label{fig:workflow}
\end{figure}

\squishlist
\item  We propose \sys{}, a pre-trained language model that learns a generic representation from multi-modal inputs, including WebAssembly code, source code, and documentation. 
This generalizable representation can facilitate understanding of low-level languages like WebAssembly and can be effectively transferred to diverse WebAssembly reverse engineering tasks.
To the best of our knowledge, this is the first multi-modal pre-trained language model for generalized WebAssembly reverse engineering.

\item Enabled by \sys{}, we develop an ML-based WebAssembly type recovery tool and a function purpose identification tool, which demand much less labeled data. In addition, we develop the first WebAssembly summarization tool. Together, these tools deliver accurate analysis results, empowering programmers with insights needed for thorough WebAssembly comprehension and further code inspection.

\item We perform a comprehensive experimental evaluation on our WebAssembly reverse engineering tools. The experiment results reveal that \sys{} achieves high accuracy and data efficiency across all tasks, surpassing state-of-the-art (SOTA) ML methods for WebAssembly.

\squishend

\section{Background and Motivation}\label{sec:motivation}



We use an motivating example to discuss the challenges in WebAssembly understanding, the limitations of state-of-the-art methods, and how our design tackles these challenges and limitations. 


\subsection{WebAssembly Basics}\label{sec:wasm_basics}
\begin{figure*}[ht!]
  \centering
  \includegraphics[width=\linewidth]{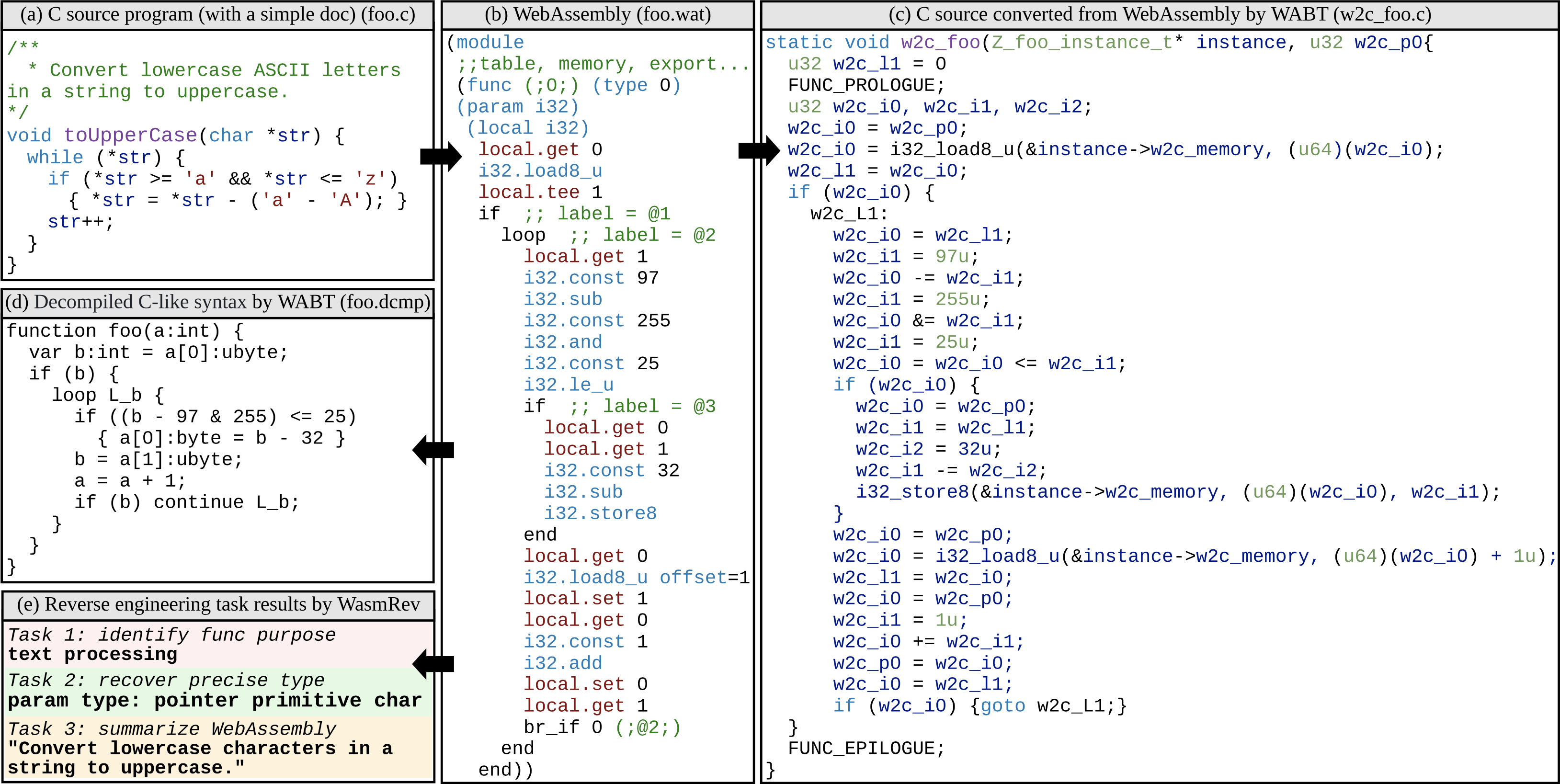}
  \caption{Motivating example. (a) A function's source code. (b) The compiled WebAssembly in text format. (c) The C source code converted from WebAssembly. (d) The decompiled C-like syntax from WebAssembly. (e) The reverse engineering task results of \sys{}. }
  \label{fig:motivationexample}
\vspace{-10pt}
\end{figure*}

The WebAssembly standard~\cite{WebAssembly_website} defines assembly-like bytecode with a unique instruction set architecture (ISA) and specific binary encodings for various types of operations. A WebAssembly binary is a binary encoding of a module with a clear modular structure composed of several sections, such as parameters and result types, functions, and data. To ensure readability, the standard provides a text format that offers a readable representation of the internal structure of a module, including type, memory, and function definitions. 
WebAssembly programs are commonly compiled from 
high-level programming languages (e.g., C, C++, and Rust) using WebAssembly compilers, such as Emscripten~\cite{Emscripten} and Rustc~\cite{rustc}. Figure~\ref{fig:motivationexample} (a) and (b) show examples of a C source code and its corresponding compiled WebAssembly in text format.

\vspace{-10pt}
\subsection{Challenges in WebAssembly Understanding}\label{sec:wasm_understand}
While WebAssembly offers promising performance and portability, the ecosystem of its reverse engineering tools is still in an early stages of development.
Substantial challenges remain for WebAssembly programmers, particularly those who are novices. 

\vspace{2pt}
\noindent\textbf{Motivating example.} 
When the high-level source code is accessible (Figure~\ref{fig:motivationexample}(a)), a programmer can comprehend the corresponding WebAssembly code (Figure~\ref{fig:motivationexample}(b)) by cross-referencing it with the relevant high-level code snippet and documentation; debugging tools such as DWARF~\cite{dwarf} may also facilitate the comprehension.
However, a common scenario often involves a WebAssembly binary delivered from a third-party, with no access to the high-level source code, source map, or debugging information. 
As a result, programmers will need to be familiar with the WebAssembly specifications and manually deduce the code behavior; alternatively, programmers may seek assistance from reverse engineering tools, such as WABT~\cite{wabt}, to convert the binary into more readable high-level code (Figure~\ref{fig:motivationexample}(c)) or decompile it into 
high-level syntax (Figure~\ref{fig:motivationexample}(d)).

We identify two major challenges with the current approaches to understanding WebAssembly.
First, \textbf{(C1) a lack of high-level information} (Figure~\ref{fig:motivationexample}(b)(c)(d)). High-level programming languages (Figure~\ref{fig:motivationexample}(a)) carry informative descriptions and convey a program's intent via function names, variable names, and natural language comments and documentation.
But WebAssembly code purely consists of low-level instructions, operating at a low level of abstraction and depicting stacking behavior. Moreover, WebAssembly only employs four data types, \textsf{i32}, \textsf{i64}, \textsf{f32}, and \textsf{f64}, significantly concealing the data type information. For instance, an \textsf{i32} type may represent a signed or unsigned integer, an array, a pointer to a struct, or many other source types. In our example, neither the converted C source (Figure~\ref{fig:motivationexample}(c)) nor the decompiled code (Figure~\ref{fig:motivationexample}(d)) successfully reconstructs lexical information or precise types:  Figure~\ref{fig:motivationexample}(c) only recovers \textsf{u32} or \textsf{u64}, which is uninformative;  Figure~\ref{fig:motivationexample}(d) recovers the input parameter type as \textsf{int}, which is ambiguous because the parameter is sometimes treated as a pointer or an array within the function body, e.g., \textsf{a[0]}. 
These gaps in representation present hurdles to precise
code reasoning and inspection. 
Second, \textbf{(C2) tracking stack behavior is cumbersome and error-prone.} WebAssembly execution is based on a stack machine architecture, which requires carefully deductive reasoning to track stack behavior. The absence of detailed debugging information and the presence of unknown optimizations substantially complicate the task of comprehending and interpreting WebAssembly. This is also less intuitive for developers accustomed to the widely adopted register-based native architectures.
In our example, the converted C source (Figure~\ref{fig:motivationexample}(c)) introduces intermediate variables and simulates the stack behavior with them using the high-level C code. However, this still requires programmers to trace the stack behavior by monitoring these variables, a process that is both tedious and error-prone, especially for non-experts.
Furthermore, the complexity of stack tracing increases with program complexity, rendering manual inspection of real-world projects increasingly impractical.

To overcome these challenges, it is imperative to furnish programmers with higher-level information, such as high-level types, function purposes, and code summarization. Such information is particularly crucial for inspecting stripped or minified WebAssembly modules that lack high-level semantics.
This higher-level information, coupled with the decompiled 
high-level syntax or the converted high-level source code,
will offer more comprehensive insights for reverse engineering, help programmers understand WebAssembly more effectively, and reduce the steep learning curve for novices.

\vspace{-8pt}
\subsection{Limitations in the State-of-the-Art Methods}\label{sec:limitOfSOTA}
\noindent\textbf{Conventional reverse engineering.} Conventional methods~\cite{brito2022wasmati,jeong2018watt,fu2018taintassembly,szanto2018taint,wabt} employ logical reasoning and rely on heuristics for WebAssembly analysis. 
The heuristics are hand-coded by experts, requiring non-trivial development effort to create, maintain, and update, especially considering the continuous evolution of WebAssembly and compilers.
Furthermore, logic-based methods are unsuitable for tasks such as code summarization, which involves uncertain information (e.g., natural language).

\vspace{2pt}
\noindent\textbf{ML-based WebAssembly reverse engineering.} Machine learning solutions, on the contrary, automatically learn heuristics and underlying features from large datasets of code and are good at handling fuzzy information~\cite{NeuralSE}. Given the promising capabilities of ML models, researchers have explored task-specific, supervised learning solutions for WebAssembly~\cite{lehmann2022finding,romano2023automated}, involving collecting a task-specific labeled dataset and supervised training a task-specific model. 
Despite the promising task performance, these studies impose several limitations. 
\textbf{(L1) Requiring large-scale labeled datasets:} Most resources are unlabeled. Despite the increasing number of open-source repositories in collaborative developer platforms like GitHub~\cite{github, markovtsev2018public}, most of the collected WebAssembly snippets are unlabeled, which can not be directly used by current supervised ML approaches. The inherent scarcity of labeled data for many WebAssembly analysis tasks necessitates extensive manual efforts for data collection, e.g., manually inspecting and labeling the purpose of functions~\cite{romano2023automated}. 
\textbf{(L2) Lacking generalization or transferability:}  task-specific models necessitating not only task-specific labeled dataset, but also customizing, training and maintenance of each model for each specific task, which is inefficient and makes them difficult to generalize or transfer
to other WebAssembly tasks. 
\textbf{(L3) Overlooking available high-level semantics:} Existing ML methods only utilize  high-level information such as high-level types and function purpose as labels in supervised learning. However, they are overlooking additional potential information and context in the high-level semantics. For example, function names and comments could provide clues about the function's intent, while the way a data object is being used in context could hint at the variable type.


\subsection{Our Design: \sys{} }

To address \textbf{(L1)}, inspired by the success of self-supervised pre-training methods in natural language processing (NLP) and software engineering~\cite{devlin2018bert,lan2019albert,feng2020codebert,wang2022jtrans}, we train \sys{} to learn a generalizable WebAssembly representation through self-supervised learning. The essence of self-supervised pre-training lies in generating ``pseudo-labels'' without the need for manual labeling. This can be easily achieved by designing ``pseudo-tasks'', for example, intentionally corrupting inputs and then training the model to identify or reconstruct these alterations. 
To resolve \textbf{(C2)}, we tailor ``pseudo-tasks'' to instruct the model to predict missing words in randomly masked WebAssembly code by leveraging context and to determine whether the order of two WebAssembly instructions has been changed (Section~\ref{sec:pre-training}, \textbf{T1} and \textbf{T3}). Through these tasks, \sys{} automatically grasps the contextual relationships, learns about control flow and stack behaviour in WebAssembly. 
The output of pre-training is a generic embedding function, which captures essential common features of WebAssembly, potentially applicable across a range of WebAssembly tasks. 
The knowledge acquired from pre-training can be generalized to various downstream reverse engineering applications through efficient few-shot fine-tuning~\cite{kadam2020review,pourpanah2022review}. This reduces the need and training time to build entirely new models from scratch, as well as the labor effort to collect large amounts of labeled data, effectively addressing limitations \textbf{(L1)} and \textbf{(L2)}.

To tackle \textbf{(C1)} and \textbf{(L3)}, we propose to learn representations of code documentation, source code and WebAssembly code concurrently -- (NL, PL, Wasm) multi-modal learning. Specifically, we design a pre-training task that guides the model to predict missing words (which can be in NL, PL, Wasm) in randomly masked multi-modal inputs, facilitating the model in learning cross-modal relationships and dependencies (Section~\ref{sec:pre-training} \textbf{T1}). 
We design another task to identify similar semantics, i.e., various modalities of the same program while distinguishing different semantics, so as to bridge the gaps among different modal representations. 
Developing such multi-model representation learning is non-trivial, due to the different language specifications and grammars. 
WebAssembly employs rigid instructions to represent stack behaviour, with instruction order significantly influenced by the compiler and optimization levels, thereby increasing contextual complexity. 
In contrast, source code employs high-level syntax and follows specific coding conventions. Natural language is weakly structured and varies in style among programmers or documentation creators. 
These factors collectively contribute to the complexity of processing multiple languages within a single neural architecture. 
Despite the success of language modeling in source code languages~\cite{feng2020codebert,guo2020graphcodebert} and native binaries~\cite{wang2022jtrans,li2021palmtree}, given the language differences, previous solutions are not ideally suitable or directly applicable to WebAssembly. To the best of our knowledge, no prior attempt has been made to learn a multi-modal representation of (NL, PL, Wasm) for WebAssembly reverse engineering. 

\vspace{-5pt}
\section{\sys{} Overview}\label{sec:overview}
To address the challenges and limitations, we propose \sys{}, the first multi-modal language model for WebAssembly understanding and reverse engineering. The goal of \sys{} is to learn a generalizable representation, i.e., learn to project multi-modal inputs into a representative embedding space, which can be used effectively in various WebAssembly reverse engineering tasks.




\vspace{2pt}
\noindent \textbf{\sys{} in a Nutshell.} As depicted in Figure~\ref{fig:workflow}, \sys{} consists of three stages: 
\squishlist
\item (1) Multi-modal pre-training (Section~\ref{sec:pre-training}): \sys{} is trained with tailored multi-modal pre-training tasks on our collected (code documentation, source code, WebAssembly code) samples, to learn a generic multi-modal embedding. 
\item (2) Few-shot fine-tuning (Section~\ref{sec:fine-tuning}): The pre-trained \sys{} is allowed for re-purposing it to various WebAssembly reverse engineering tasks. To evaluate the generalization of the pre-trained \sys{}, we fine-tune it on three tasks: type recovery, function purpose identification and WebAssembly summarization. The fine-tuning is few-shot, requiring much less task-specific labeled datasets than non pre-train methods. 
\item (3) Inference: Each of the fine-tuned models performs inference on new WebAssembly input, and generates accurate prediction. The collected outputs together serve programmers with high-quality insights who decide to perform further code inspection. An example report is shown in Figure~\ref{fig:motivationexample} (e).
\squishend

\section{Learn a generic multi-modal 
Representation}\label{sec:pre-processing}
\begin{figure}[h]
\centering
  \includegraphics[width=\linewidth]{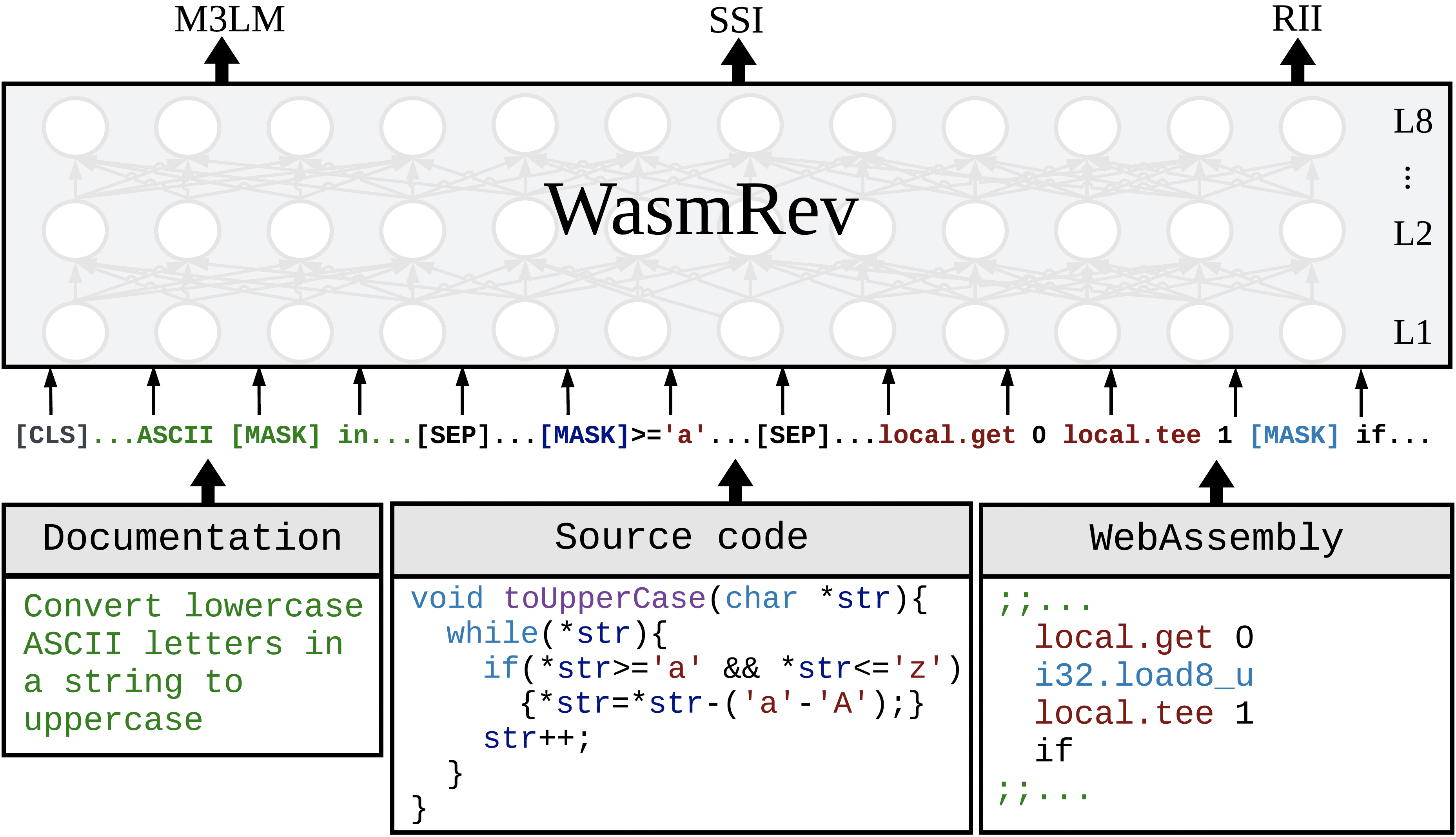}
  \caption{An illustration of \sys{} pre-training. The model architecture consists of 8 bidirectional Transformer layers. The model takes the concatenation of (documentation, source code, WebAssembly code) as input, and is pre-trained on M3LM, SSI and RII tasks.}
  \vspace{-10pt}
  \label{fig:model}

\end{figure}

In this section, we will elaborate how \sys{} learns a generic (NL, PL, Wasm) multi-modal representation. 
As illustrated in Figure~\ref{fig:model}, \sys{} comprises three key components: \sys{} model, input/output representations and pre-training tasks.

\subsection{\sys{} Model Architecture}
We adopt BERT~\cite{devlin2018bert} as \sys{}'s backbone, inspired by its promising performance in various code-related tasks~\cite{feng2020codebert,guo2020graphcodebert}. As the model architecture is standard, we will not review the Transformer architecture in detail. As will be elaborated later, we adapt the model through our tailored pre-training process to be \sys{} for better  WebAssembly comprehension. 

\subsection{Input / Output Representations} Previous WebAssembly ML models~\cite{lehmann2022finding,romano2023automated} consider only WebAssembly as input. In this paper, we argue that aggregating complementary information from semantically equivalent modalities -- source code, code documentation and WebAssembly code -- is benefit for WebAssembly comprehension, potentially better supporting diverse downstream tasks and yielding promising results (Section~\ref{sec:effectiveness}). 
Given a NL documentation $c = \{c_1, c_2, . . . , c_{|c|}\}$, we consider it as a sequence of words, and split it as WordPiece~\cite{wu2016google}. Given the corresponding source code in a high level PL $s = \{s_1, s_2, . . . , s_{|s|}\}$, and the compiled WebAssembly $w = \{w_1, w_2, . . . , w_{|w|}\}$, we regard them as sequences of tokens. \sys{} takes the concatenation of multiple modalities (NL, PL, Wasm) as model input, i.e.,
\begin{equation}
\small
    x = \{[CLS], c, [SEP], s, [SEP], w, [SEP]\}
\end{equation}
\normalsize

\noindent where [CLS] is a special token at the beginning of the input sequence, whose final hidden representation is considered as the aggregated sequence representation for classification or ranking; and [SEP] is a special token to split two kinds of sub-sequences~\cite{devlin2018bert}. We normalize instructions to avoid the out-of-vocabulary problem, following previous studies on native binary representation learning~\cite{wang2022jtrans,li2021palmtree}. Specifically, string literals are replaced by a special token [STR]. Large constants, empirically those greater than 0xffff, which likely represent memory addresses, are normalized by a special token [ADDR]. Small constants are retained as individual tokens as they may carry valuable information, such as specifying which local variables or function arguments to access. 
Then we build a vocabulary upon the normalized corpus and tokenized the inputs. We also incorporate position embedding and segmentation embedding into token embedding, and use the mixed vector of them as model input. Specifically, position embedding represents different positions in the input sequence, while segmentation embedding distinguishes documentation, source code and WebAssembly code. 

The output of \sys{} includes (1) contextual vector representation of each token, for code documentation, source code, and WebAssembly code; (2) the representation of [CLS], which works as the aggregated sequence representation.

\subsection{\textbf{Pre-training Tasks}} \label{sec:pre-training}
To best adapt \sys{} for multi-modal learning, particularly for WebAssembly reverse engineering, we go beyond simply adopting BERT's original pre-training approach. 
Instead, we design three self-supervised pre-training tasks that are specifically crafted to exploit and learn from our multi-modal input representation, enabling \sys{} to grasp inter-modal and intra-modal relationships. 

\begin{figure}[h]
  \centering
  \includegraphics[width=\linewidth]{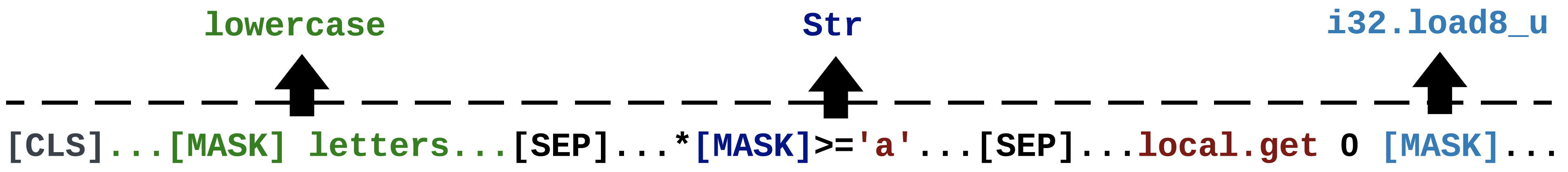}
  \caption{A multi-modal masked language model example}
  \vspace{-5pt}
  \label{fig:cmlm}

\end{figure}

\vspace{3pt}
\noindent\textbf{(T1) Multi-Modal Masked Language Model (M3LM)}
Our approach jointly models NL, PL, and Wasm, providing complementary information contained in multiple modalities. We first extend the NLP task Masked Language Model (MLM) of BERT to multiple modalities. Given a data point of (NL, PL, Wasm) triplet $\{c, s, w\}$ as input, we randomly select $15\%$ of tokens from the concatenation of (NL, PL, Wasm). We show an M3MLM example in Figure~\ref{fig:cmlm}. For the selected tokens, we replace $80\%$ of them with [MASK] tokens, $10\%$ with random tokens, and the remaining $10\%$ is unchanged. Formally, the loss function of M3LM is defined as:

\begin{equation}
\small
    \mathcal{L}_{M3LM} = -\sum_{i \in M} log P(t_i|f_M)
\end{equation}

\normalsize

\noindent where $M$ is the set of indices of the masked or replaced tokens, and $f_M$ is the corrupted input, $P(t_i|f_M)$ is the probability for predicting a particular token $t_i$, with $f_M$ as input, calculated by the representation of the token from \sys{} following a softmax function to normalize the output. 

By extending conventional MLM to M3LM, the model is guided to learn not only the intra-modal contextual relationship to reconstruct missing words, but also the cross-modal dependencies. In particular, if the WebAssembly context alone proves insufficient for deducing the masked WebAssembly token, the model can draw upon both documentation and source code to enhance its understanding. This also holds true for the masked token prediction for both documentation and source code.

\noindent\textbf{(T2) Similar Semantics Identification (SSI)}
To bridge the gap among different modalities, we task the model with identifying similar semantics, i.e., various modalities of the same program. 
While the documentation, source code and WebAssembly code of the same program correspond to the same program semantics, they are dissimilar to representations of another program. 
As such, we create semantically similar and dissimilar pairs as model inputs. 
There are numerous ways to create these semantically similar pairs, including single-modal pairs like NL \textit{vs.} Wasm, bi-modal pairs like (NL, PL) \textit{vs.} (NL, Wasm), and triple-modal pairs like (NL, PL, Wasm) \textit{vs.} (NL, Wasm, PL). 
Our selection of semantic-similar pairs is guided by potential downstream applications.
Motivated by the WebAssembly summarization task, a Wasm-NL task, we include NL \textit{vs.} Wasm. Considering the potential WebAssembly decompilation tasks, which are Wasm-PL tasks, we incorporate PL \textit{vs.} Wasm. With the aim of enriching the variety of similar pair formats to enhance the SSI task, we further include additional bi-modal and triple-modal pairs. 
We include (NL, PL) \textit{vs.} (NL, Wasm) pairs, which encourage the model to learn the semantics of PL and Wasm, with NL providing context. Similarly, we include (NL, PL) \textit{vs.} (PL, Wasm) to enable our model to learn the semantics behind NL and Wasm, with PL as context. We also include (NL, PL, Wasm) \textit{vs.} (NL, Wasm, PL) to help the model to learn semantics behind different orders of modalities. 
Furthermore, the samples with WebAssembly code compiled from the same source code but with different compilation flags are also semantically similar pairs.

\begin{figure}[h]
  \centering
  \includegraphics[width=0.88\linewidth]{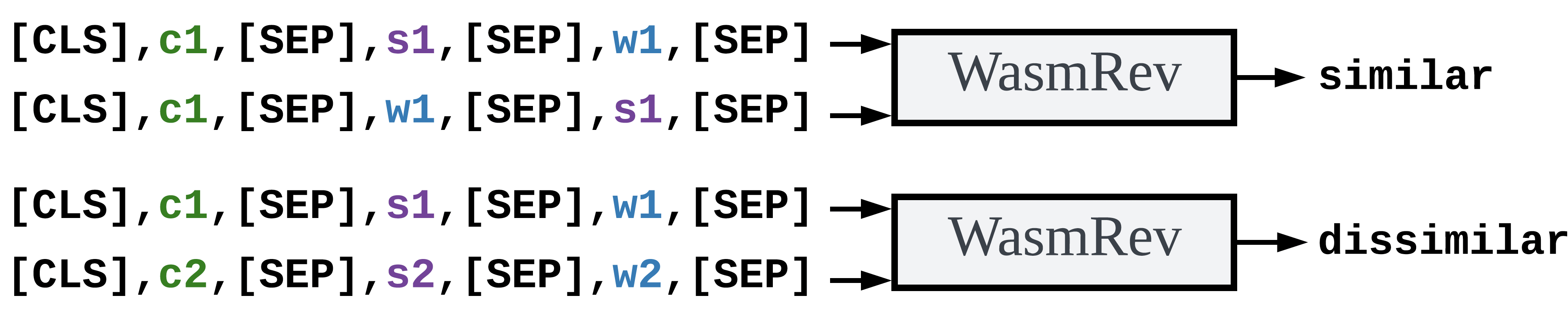}
  \caption{A similar semantics identification examples}
\vspace{-5pt}
  \label{fig:ssi}
\end{figure}

We then employ in mini-batch and cross mini-batch sampling strategies~\cite{chen2020simple} to construct dissimilar samples. For a batch of training data $b_1 = [x_1 ...x_N]$ with batch size $N$, we can first obtain a batch of data $b_2 = [x^+_1 ...x^+_N]$ where $(x_i , x^+_i)$ is a pair of similar semantics as describe above, e.g., NL \textit{vs.} Wasm. 
The dissimilar semantics $x_i^-$ of $x_i$ is chosen as $x_j$ and $x_j^+$, $\forall i \neq j$ and $x_i$ and  $x_j/x_j^+$ are corresponding to different source code. We show an example of SSI in Figure~\ref{fig:ssi}. For an input $x_i$ with representation $v_i$, the goal of SSI is to maximize the representation similarity (dot product) between similar samples, while minimizing the representation similarity between dissimilar samples. Formally, the loss of SSI is defined as:
\begin{equation}
\small
    \mathcal{L}_{SSI} = -\sum_{i\in \{\mathcal{N},\mathcal{N^+}\}}ln \frac{exp(v_i \cdot v_i^+)}{exp(v_i \cdot v_i^+)+\sum_{k=1}^{|x_i^-|}exp(v_i \cdot v_k^-)}
\end{equation}
\normalsize

\noindent where $\mathcal{N}$ and $\mathcal{N^+}$ are the sets of all program samples covering similar samples discussed above.

\begin{figure}[h]
  \centering
  \includegraphics[width=0.8\linewidth]{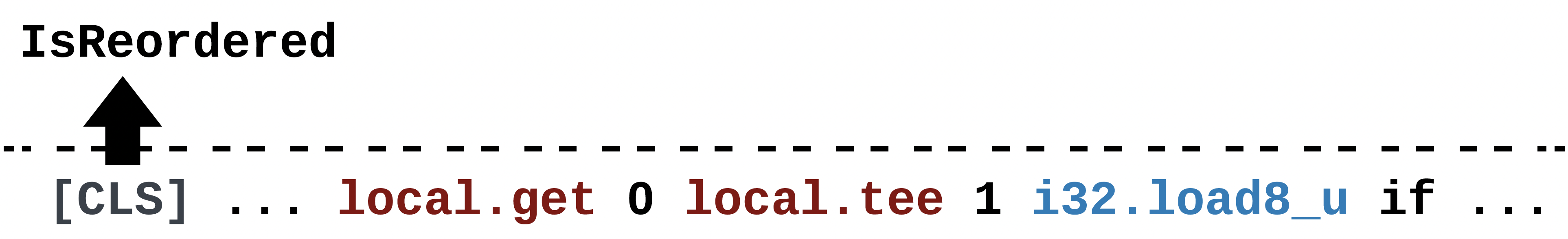}
  \caption{An reordered instructions identification example}
\vspace{-5pt}
  \label{fig:iop}
\end{figure}

\noindent\textbf{(T3) Reordered Instructions Identification (RII)}
Since our target is WebAssembly understanding, we design a task specifically aimed at enabling the model to learn WebAssembly representation, especially the contextual and control flow features. 
In WebAssembly, the control flow is well-structured. Due to the stack-based architecture of WebAssembly, each instruction controls the next instruction in code. 
As such, we design a reordered instructions identification task to help \sys{} comprehend the control flow relationship and the intended sequence of information in code. 
Specifically, we randomly sample 20\% of consecutive instruction pairs for example $(I_1,I_2)$ and swap its sequence to be $(I_2,I_1)$, and then ask \sys{} to predict whether the instructions are reordered or not. We show an example of RII in Figure~\ref{fig:iop}, where the instructions ``\textsf{local.tee 1}'' and ``\textsf{i32.load8\_u}'' are swapped. The loss function of RII objective is defined as:

\begin{equation}
\small
   \mathcal{L}_{RRI} = 
   -\sum_{I_1,I_2\in I} [y_{1,2}logP(y_{1,2}|I_1,I_2)) + (1-y_{1,2})log(1-P(y_{I_1,I_2}|I_1,I_2)))]
\end{equation}
\normalsize

\noindent where $I$ is a set of instruction candidates for RRI, ($I_1, I_2$) is an instruction pair as input.  
The probability $P(y_{1,2}|I_1,I_2)$ of instruction reordering is derived from the representation of the \textsf{[CLS]} token from \sys{}, following a softmax function.


Finally the pre-training objective is to optimize \sys{} to minimize the loss function $\mathcal{L}$ combining three tasks' loss functions:

\begin{equation}
\small
     \mathcal{L} = \mathcal{L}_{M3MLM}+\mathcal{L}_{SSI}+\mathcal{L}_{RII}+\lambda||W||^2
\end{equation}
\normalsize

where $W$ contains all trainable parameters of the model. $\lambda$ is the coefficient of $L_2$ regularizer to prevent overfitting.

\section{Developing WebAssembly Reverse Engineering Tools} \label{sec:fine-tuning}

By exploiting the generalization of \sys{} and knowledge acquired from pre-training, we develop three WebAssembly reverse engineering tools for function purpose identification (FPI), type recovery (TR) and WebAssembly summarization (WS) by fine-tuning the pre-trained \sys{} correspondingly. These reverse engineering tasks take only the WebAssembly bytecode as input to generate high-level semantics, such as high-level types and function purpose, thereby facilitating the comprehension of WebAssembly programs.

\subsection{\textbf{Function Purpose Identification (FPI)}}\label{sec:fpi}

Function purpose identification is a task to identify the purpose of a given WebAssembly function, e.g., numeric processing, cryptography, etc. It provides developers with the type of a function, aiding in code understanding and further code inspection. We use the same problem setting as described in WASPur~\cite{romano2023automated} and formulate FPI as a classification task to identify the type of a function from  a set of predefined types. 
By leveraging the learnt WebAssembly representation in pre-training, we fine-tune \sys{} to accurately predict function types. Specifically, we use the last hidden state of [CLS] token for classification to fine-tune \sys{}.

\subsection{\textbf{Type Recovery (TR)}}\label{sec:tr}
Type recovery is a task to generate high-level precise types for parameters and return value of a function. It is an important first step toward WebAssembly comprehension, and is widely explored in existing reverse engineering tools for native binaries~\cite{caballero2016type,pei2021stateformer,chua2017neural}. 
We employ the same problem setting as described in SnowWhite~\cite{lehmann2022finding} and treat the type recovery problem as a sequence prediction task, where the sequence comprises tokens in a  high-level type language. 
We leverage the knowledge of source code-WebAssembly relationship learnt in pre-training, and fine-tune \sys{} to recover parameters and return types. We employ an encoder-decoder model for this Wasm-to-PL generation task. Specifically, we use the pre-trained \sys{} to initialize the encoder and use a simple 2-layer bi-LSTM with an attention mechanism as the decoder. 

\subsection{\textbf{WebAssembly Summarization (WS)}}\label{sec:ws}
Code documentation is a crucial task to bridge the gap between complex code and understandable documentation, facilitating easier code comprehension, maintenance, and collaboration. Code documentation has been widely studied for source code~\cite{feng2020codebert,alon2018code2seq,lu2021codexglue}. Yet, the specific task of WebAssembly summarization has received limited attention. Existing tools like WASPur, which predicts the general purpose of functions, offering a rudimentary form of summarization that can be treated as a preliminary summarization. However, it falls short in capturing the specific functionalities of the code. For instance, WASPur might broadly classifies a function as `numerical processing' rather than precisely stating it `return the sum of two numbers.', necessitating further code inspection for concrete understanding. To the best of our knowledge, this is the first work to formally define the WebAssembly summarization and propose a ML solution for it. We define WebAssembly summarization as a sequence generation task to generate human-readable natural language descriptions or summaries of a given WebAssembly code, i.e.,  what a particular piece of WebAssembly code does. This requires the model to understand not just the syntax but also the semantics of the WebAssembly code. A good summarization should be both factually correct, and useful and readable to the end-user. For WS, since it is a Wasm-to-NL generation task, we employ the same encoder-decoder architecture as utilized in the TR task, and fine-tune it on WS dataset.

\section{Implementation}\label{sec:implementation}
\subsection{\textbf{Pre-training Dataset}}\label{sec:pre-train-dataset}
To facilitate 
generalizability of \sys{} and effective representation learning in pre-training, we collect a large scale of C/C++ code paired with documentation, from 432 publicly available open-source non-fork GitHub repositories, and prefer the high-quality projects indicated by their number of stars and forks. 
To obtain a representative dataset that has a close distribution to the realistic WebAssembly distribution, we follow the common WebAssembly use cases~\cite{wasm_use_cases} to collect a broad spectrum of applications spanning diverse domains, e.g., gaming, text/media/numerical processing, machine learning, cryptography, etc., and covering varying complexities and code styles. 

We remove duplicates from the dataset by identifying (near) duplicate functions and only keeping one copy of them, following the de-duplication method proposed by Allamanis~\cite{allamanis2019adverse}. 
For the documentation, we truncate it to the first full paragraph, which typically summarizes and describes a function, while removing in-depth discussion of function arguments and return values. 
Samples with documentations shorter than three tokens are filtered out, as they are not informative. 
After preprocessing and filtering the source code and documentation, we collect 378k C/C++ functions paired with their documentation. 
We allocate only 80\% of them as pre-training data of \sys{}, while the remaining 20\% (75.6k) are used for the WS task, ensuring that \sys{} has never seen the source code and documentation in the WS task. 
We then compile the source code using Emscripten~\cite{Emscripten} v3.1.12 with different optimization options \textsf{-O0}, \textsf{-O1}, \textsf{-O2}, \textsf{-O3}, \textsf{-Os}, \textsf{-Oz}, allowing \sys{} to learn WebAssembly code with various compilations. We employ WABT~\cite{wabt} to transform WebAssembly into the text format. During this process, invalid samples that cannot be compiled by Emscripten or transformed by WABT are filtered out. For code file consisting of multiple functions, we first obtain the unstripped binary to identify the offsets of functions, and then use the offsets to locate the functions in the striped binary and label them with the corresponding documentation. 
Ultimately, we collect $\sim1.8M$ (code documentation, source code, WebAssembly code) samples for \sys{} pre-training. 

\subsection{\textbf{Fine-tuning Datasets}}\label{sec:fine-tuning-dataset}
\subsubsection{FPI dataset:} \label{sec:fpi-dataset}
We use the dataset from WASPur~\cite{romano2023automated}, consisting of 1,829 WebAssembly files collected from real-world websites, Firefox add-ons, Chrome extensions, and GitHub repositories, with 151,662 functions manually classified into 12 categories (listed in  Figure~\ref{fig:tsne}). 
\subsubsection{TR dataset:} \label{sec:tr-dataset}
We use the dataset from SnowWhite~\cite{lehmann2022finding}, composed of 5.5 million parameter type samples and 796 thousand return type samples, labeled with high-level types in its proposed expressive type language. It should be noted that the utilization of source code containing high-level types in C/C++ in pre-training does not result in data leakage for the TR task. This is because the TR task employs the type language defined in SnowWhite, making the targeted type sequences distinct from the C/C++ types used in pre-training data.
\subsubsection{WS dataset:} \label{sec:ws-dataset}
As mentioned in Section~\ref{sec:pre-train-dataset}, we use 75.6k C/C++ code with documentation labels (which the pre-trained \sys{} has never seen) for WS task, and compile them with Emscripten and six various optimization options. We finally have 453k (WebAssembly code, documentation) samples for WS task.

All the datasets for fine-tuning tasks are split by 80\%: 10\% :10\% into training / validation / testing sets. For data splitting, we split the source code first, then compile them using different optimization settings. In this way, we ensure functions from one project are not split over training, testing, and validation sets.

\subsection{\sys{} Implementation} 
We implement our \sys{} model using PyTorch v1.11~\cite{collobert2002torch} in python. For \sys{}'s neural architecture, we apply 8 Transformer layers, with 128 hidden states, 8 attention heads, initialized with Xavier~\cite{glorot2010understanding}. We set the max sequence length as 512 and use the Adam optimizer~\cite{Adam} with a learning rate of 0.0005, $\beta_1$ = 0.9, $\beta_2$ = 0.999, weight decay rate of 0.01, and linear decay of the learning rate. To mitigate the over-fitting problem, we employ a dropout technique with a dropping ratio of 0.1. We pre-train \sys{} model with a batch size of 32 for 5 epochs using the pre-training dataset as described in Section~\ref{sec:pre-train-dataset}. Subsequently, we fine-tune it with the same batch size for 2 epochs on each downstream task and set the learning rate to 3e-5, utilizing the fine-tuning dataset outlined in Section~\ref{sec:fine-tuning-dataset}. During the fine-tuning phase for each downstream task, we fine-tune the pre-trained \sys{} model using the corresponding training set, and monitor accuracy on the validation set to facilitate early stopping upon model convergence. The best-performing model from the validation phase is then selected for the final evaluation using the testing set. All final predictions are obtained on the test data, which the model has never seen and was not used to select the best model. 
All the experiments are run on a Linux server running Ubuntu 20.04 with Intel Xeon E5-2686 v4 CPU with 12 cores running at 2.3Ghz, 488GB memory, and 8 Nvidia V100 16G GPUs.

\section{Evaluation}\label{sec:evaluation}




In this section, we report and analyze the experimental results to answer the following research questions (RQ):
\squishlist
\item \textbf{RQ1:} Can \sys{} accurately distinguish different semantics of WebAssembly? 
\item \textbf{RQ2:} Can \sys{} accurately learn the relationship of WebAssembly and high-level programming language? 
\item \textbf{RQ3:} Can \sys{} accurately grasp the relationship of WebAssembly and natural language? 
\item \textbf{RQ4:} How effective are the multiple modalities of input representations?
\item \textbf{RQ5:} How effective are our designed pre-training tasks?
\item \textbf{RQ6:} How is the data-efficiency of \sys{} fine-tuning?
\squishend

\subsection{\textbf{Function Purpose Identification (RQ1)}}
To study whether \sys{} can accurately distinguish between the various semantics of WebAssembly (RQ1), we fine-tune the pre-trained \sys{} model and evaluate it on the function purpose identification (FPI) task as described in Section~\ref{sec:fpi}, using the fine-tuning dataset specified in Section~\ref{sec:fpi-dataset}. FPI is a WebAssembly classification task, where the input is a WebAssembly function and output is the purpose of the input function, e.g., numeric processing, cryptography. The FPI task evaluates \sys{}'s ability on understanding and recognizing various semantic patterns of WebAssembly functions by requiring it to categorize these functions based on their specific purposes. 

\noindent{\textbf{$\bullet$ Baseline.}} We use the SOTA model WASPur as our baseline for FPI, which 
employs a DNN model with an embedding layer, three hidden layers 
and an output layer, with supervised learning.

\noindent{\textbf{$\bullet$ Metrics.}} We adopt four metrics used in baseline WASPur: accuracy (Acc), precision (P), recall (R), F1 score. 
They are calculated by true positive (TP), true negative (TN), false positive (FN) and false negative (FN), where true or false identify whether the prediction is correct or not, and positive and negative indicate the positive class or negative class. The metrics are defined as:

\begin{equation}
\footnotesize
    Acc=\frac{TP+TN}{TP+TN+FP+FN},   \ P=\frac{TP}{TP+FP},  \   R=\frac{TP}{TP+FN},  \ F1 = \frac{2*P*R}{P+R}
\end{equation}
\normalsize


\noindent{\textbf{$\bullet$ Results.}} We show the results in Table~\ref{tab:FPI}. \sys{} outperforms WASPur by 6.18\%, 3.30\%, 5.98\%, 4.61\% in terms of top-1 accuracy, precision, recall and F1 score, 
revealing the effectiveness of \sys{} to learn WebAssembly representations, and recognize diverse semantic patterns among various types of functions.

\Huge
\begin{table}[h]
\caption{FPI and WS results, and effectiveness of design components. \sys{} and its variants are fine-tuned separately for various tasks.}
\label{tab:FPI}
\resizebox{\linewidth}{!}{
\begin{threeparttable}
\begin{tabular}{ccccccccc}\toprule
Task &\multicolumn{4}{c}{FPI} &\multicolumn{1}{c}{\begin{tabular}[c]{@{}c@{}}TR \\ param\end{tabular}} & \multicolumn{1}{c}{\begin{tabular}[c]{@{}c@{}}TR \\ return\end{tabular}}& \multicolumn{2}{c}{WS} \\ \cmidrule(r){2-5} \cmidrule(r){6-7}\cmidrule(l){8-9} 
Metric &Acc\tnote{$*$}& P&R &F1&Acc\tnote{$*$}&Acc\tnote{$*$}&BLEU&BF1\tnote{$\ddag$}\\   \hline
         WASPur\tnote{$\dag$}&   $88.07\%$ & 0.91& 0.87&0.89&-&-&-& -\\ \hline 
         \sys{}\textsubscript{SuV(1/10)}(WS)&-&-&-&-&-&-&18.62&0.765\\ 
         \sys{}\textsubscript{SuV}(WS)&-&-&-&-&-&-&19.57&0.833\\ \hline
         \sys{}\textsubscript{w/oNL}&93.17\%&0.935&0.919&0.927&63.2\% &72.5\% &19.88&0.840\\ 
         \sys{}\textsubscript{w/oPL}&93.20\%&0.933&0.917&0.925&62.8\% &72.1\% &19.57&0.831\\ 
         \sys{}\textsubscript{w/oNL+PL}&92.88\%&0.921&0.910&0.915&62.5\% &71.9\%&19.49&0.828\\ \hline
         \sys{}\textsubscript{w/oM3MLM}&93.06\%&0.930&0.915&0.922&62.2\%&71.8\%&20.00&0.851\\ 
         \sys{}\textsubscript{w/oSSI}&93.00\%&0.929&0.916&0.922&62.8\%&72.2\%&20.10&0.855\\ 
         \sys{}\textsubscript{w/oRII}&92.92\%&0.925&0.908&0.916&62.6\%&72.1\%&20.12&0.855\\ \hline
         \sys{}\textsubscript{(1/10)}&91.25\%&0.920&0.901&0.910&61.9\%&70.5\%&19.32&0.831\\ 
         \sys{}& \textbf{93.51\%}& \textbf{0.940}& \textbf{0.922}&\textbf{0.931}&\textbf{63.7\%}&\textbf{72.9\%}& \textbf{20.33}&\textbf{0.873} \\  \hline\end{tabular}
\begin{tablenotes}
\item[$*$] ``Acc'' indicates top-1 accuracy.
\item[$\dag$] The best results reported in the WASPur paper.
\item[$\ddag$] ``BF1'' indicates BERTScore F1 score.
  \end{tablenotes}
\end{threeparttable}}
\vspace{-10pt}
\end{table}
\normalsize

\begin{figure}[h]
\centering
  \includegraphics[width=0.9\linewidth]{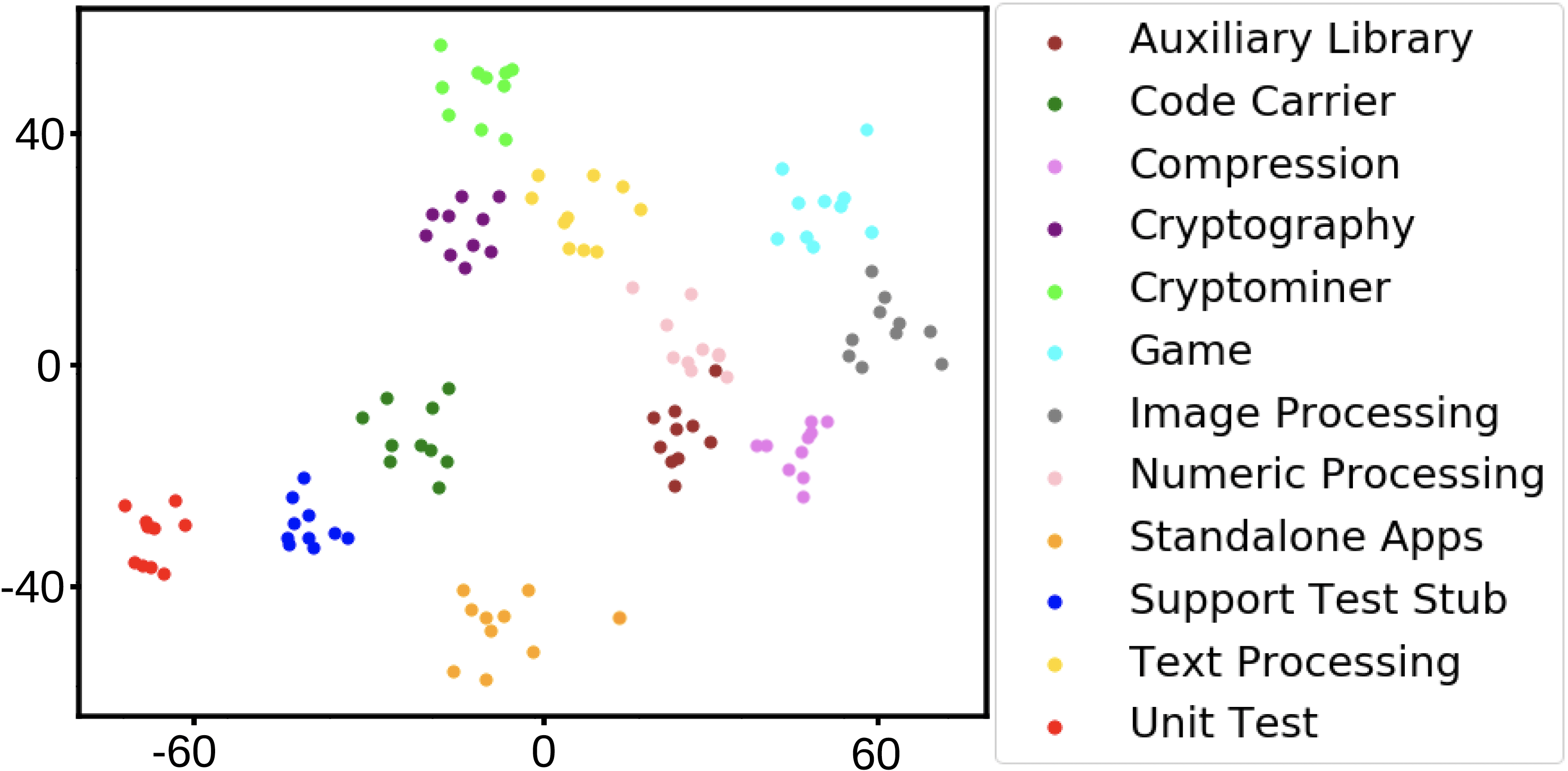}
  \captionof{figure}{Visualization of FPI results of \sys{}.}
  \label{fig:tsne}
\vspace{-5pt}
\end{figure}

\noindent{\textbf{$\bullet$ Visualization.}}
To intuitively show how well does \sys{} distinguish different semantics of WebAssembly (RQ1), we visualize the classification results. 
In particular, we randomly select 10 functions from each class of function purpose, and compute the embeddings of them. We then use the t-distributed Stochastic Neighbor Embedding (t-SNE) method~\cite{van2008visualizing} to project the high-dimensional embeddings onto a 2-D plane. We show the visualization in Figure~\ref{fig:tsne}, with different classes marked in different colors. The clear clustering of functions of the same type and the separation between functions of different types, reveal \sys{}'s capability of identifying and distinguishing various types of functions. Interestingly, we observe the representations of \textit{Standalone Apps} are relative scattering, since various Apps could have diverse functionalities. One potential improvement is to refine the dataset, e.g., divide samples in \textit{Standalone Apps} into sub-classes of similar functionalities. We also observe some clusters are close to each other, e.g.,  \textit{Numeric Processing} and  \textit{Auxiliary Library}, which could both involve utility functions, possibly with a mathematical focus. 

\begin{figure}[h]
\centering
  \includegraphics[width=\linewidth]{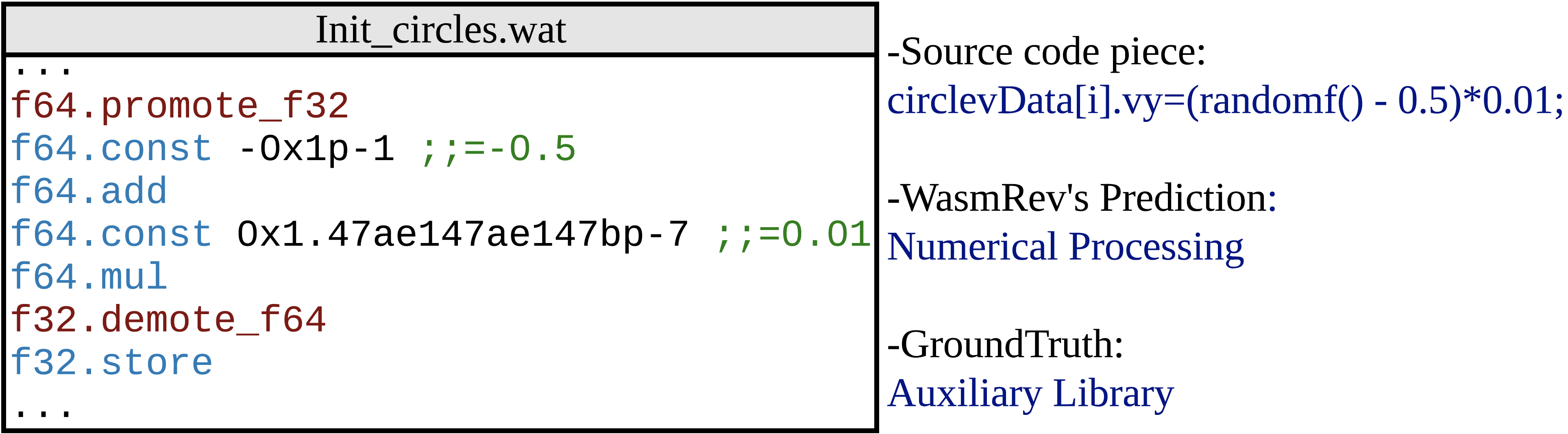}
  \captionof{figure}{A case study of FPI.}
  \label{fig:fpi_casestudy}
\vspace{-8pt}
\end{figure}

\noindent{\textbf{$\bullet$ Case Studies.}}
We show two case studies of FPI. Since we do not have access to the source code of WASPur, we limit our discussion to our results. The first case is shown in Figure~\ref{fig:motivationexample}, where \sys{} accurately identifies the function type as \textsf{text processing}. The second case, illustrated in Figure~\ref{fig:fpi_casestudy}, is a function to initialize the positions and velocities of circles to be used in a graphical application, belong to the \textit{Auxiliary Library} type. Here, \sys{} incorrectly categorizes the function type as \textit{Numerical Processing}, since the function involves a lot of mathematical operations as shown in the WebAssembly code \textsf{Init\_circles.wat}. This aligns with the observation in the visualization that the ambiguity of code and labeling could lead to incorrect predictions. To address this issue, we could use the results from the WS task to aid in better comprehending the function purpose.

\begin{tcolorbox}[
  enhanced,
  colback=blue!4!white,
  boxrule=0.8 pt, 
  boxsep=0pt, 
  left=2pt, 
right=2pt, 
  top=2pt, 
  bottom=2pt, 
]
\textbf{Result-1:} \sys{} is able to accurately distinguish different semantics of WebAssembly, with a high accuracy (93.51\%) to identify the correct class of purpose of a WebAssembly function, outperforming WASPur by 6.18\% in top-1 classification accuracy.
\end{tcolorbox}

\Huge
\begin{table*}[t!]
\caption{TR results.}
\label{tab:type}
\resizebox{0.9\linewidth}{!}{
\begin{threeparttable}
\begin{tabular}{cccccccccccc}
\toprule
&Task      & \multicolumn{5}{c}{Parameter Type Prediction}   & \multicolumn{5}{c}{Return Type Prediction}    \\\cmidrule(r){3-7} \cmidrule(l){8-12}
 &Type Language     & $\mathcal{L}_{SW}$ & \begin{tabular}[c]{@{}c@{}} $\mathcal{L}_{SW}$, \\ All Names\end{tabular} & \begin{tabular}[c]{@{}c@{}} $\mathcal{L}_{SW}$, \\ Simplified\end{tabular} &  $\mathcal{L}_{Eklavya}$ & \begin{tabular}[c]{@{}c@{}}$\mathcal{L}_{SW}, t_{low}$\\ not given\end{tabular} & $\mathcal{L}_{SW}$ & \begin{tabular}[c]{@{}c@{}}$\mathcal{L}_{SW}$, \\ All Names\end{tabular} & \begin{tabular}[c]{@{}c@{}}$\mathcal{L}_{SW}$, \\ Simplified\end{tabular} & $\mathcal{L}_{Eklavya}$ & \begin{tabular}[c]{@{}c@{}}$\mathcal{L}_{SW}, t_{low}$\\ not given\end{tabular} \\ \midrule
{\multirow{3}{*}{SnowWhite\tnote{$\dag$}}} &Top-1 Acc         &     44.5\% & 18.6\% & 65.1\% & 87.9\% & 43.4\% & 57.7\% &40.6\%&60.6\%&76.3\%&50.7\% \\
 &Top-5 Acc         &     75.2\% & 27.1\% & 86.2\% & 100.0\% & 74.3\% & 80.5\% &47.3\%&87.9\% &100\% &81.2\%\\
 &Type Prefix Score &     1.47 & 1.31 & 1.62 & 0.88 & 1.45 & 1.37&1.00&1.38&0.76&1.02  \\\midrule
{\multirow{3}{*}{\sys{}}}&Top-1 Acc         &    \textbf{ 63.7\%} & 40.2\% & 80.6\% & 93.4\% & 62.8\% & \textbf{74.9\%} &52.4\%&79.7\%&89.2\%&73.5\% \\
 &Top-5 Acc         &     88.3\% & 50.0\% & 95.2\% & 100.0\% & 87.4\% & 93.8\% &63.9\%&95.1\% &100\% &93.2\%\\
&Type Prefix Score &     1.78 & 1.55 & 1.89 & 0.93 & 1.76 & 1.60&1.30&1.58&0.89&1.31  \\\bottomrule              
\end{tabular}
\begin{tablenotes}
\item[$\dag$]The results reported in the SnowWhite paper.
  \end{tablenotes}
\end{threeparttable}}
\vspace{-10pt}
\end{table*}
\normalsize

\subsection{\textbf{Type Recovery  (RQ2)}}
To investigate whether \sys{} can accurately learn the relationship of WebAssembly and high-level programming language (RQ2), we fine-tune the pre-trained \sys{} model and evaluate it on the type recovery (TR) task as detailed in Section~\ref{sec:tr}, with the fine-tuning dataset described in Section~\ref{sec:tr-dataset}. TR is a sequence generation task, where the input is a WebAssembly function and output is a sequence in high level programming language that specifies the parameters and return types, e.g., \textsf{primitive int32}, \textsf{pointer primitive char}. TR evaluates how well \sys{} can map concepts and types in low-level WebAssembly to high-level type signatures by understanding the WebAssembly context.

\noindent{\textbf{$\bullet$ Baseline.}} 
We use the SOTA SnowWhite as our baseline, which uses supervised learning on a sequence-to-sequence model~\cite{sutskever2014sequence}.

\noindent{\textbf{$\bullet$ Metrics.}} We adopt three metrics used in baseline SnowWhite. (1) top-1 accuracy (perfect match accuracy). (2) top-5 accuracy, which retrieves the five most likely type predictions via beam search. (3) The average type prefix score (TPS) over the whole test set, where the TPS is defined as the length of the common prefix $TPS(t',t) = |commonPrefix(t',t)|$ of a prediction $t'$ and ground truth $t$. 

\noindent{\textbf{$\bullet$ Results.}}
We show the results in Table~\ref{tab:type}, with various high-level type languages provided in the SnowWhite dataset. With the $\mathcal{L}_{SW}$ type language and 1,225 unique types in the testing set, \sys{} achieves 63.7\%/88.3\% top-1/top-5 accuracy for parameter type prediction and 74.9\%/93.8\% top-1/top-5 accuracy for return type prediction, outperforming SnowWhite by 43.1\%/17.4\% and 29.8\%/16.5\% correspondingly. \sys{} achieves an average type prefix score of 1.78 for parameters and 1.60 for return values. This indicates that \sys{} can accurately predict the first 1.78 and 1.60 tokens of the type sequence for parameters and return values, respectively. The results reveal \sys{}'s effectiveness in type recovery, since we expect the one or two tokens to be likely the most relevant to a reverse engineer. 
($\mathcal{L}_{SW}$, All Names) is a variant of $\mathcal{L}_{SW}$ with non-filtered type names and 146,883 different types in total, making the TR task more difficult. ($\mathcal{L}_{SW}$, Simplified) is a simplified variant of $\mathcal{L}_{SW}$, with some constructors removed in the type language to consist of 120 types. $\mathcal{L}_{Eklavya}$ has the simplest setting with only 7 different types. ($\mathcal{L}_{SW}, t_{low}$ not given) is similar to $\mathcal{L}_{SW}$, but removes the low-level type information in the input. Among all these settings, \sys{} achieves better results than SnowWhite across all the metrics, demonstrating its effectiveness in understanding the relationship between WebAssembly and high-level programming language, and projecting the correct types from WebAssembly. 



\begin{figure}[t]
\centering
  \includegraphics[width=0.75\linewidth]{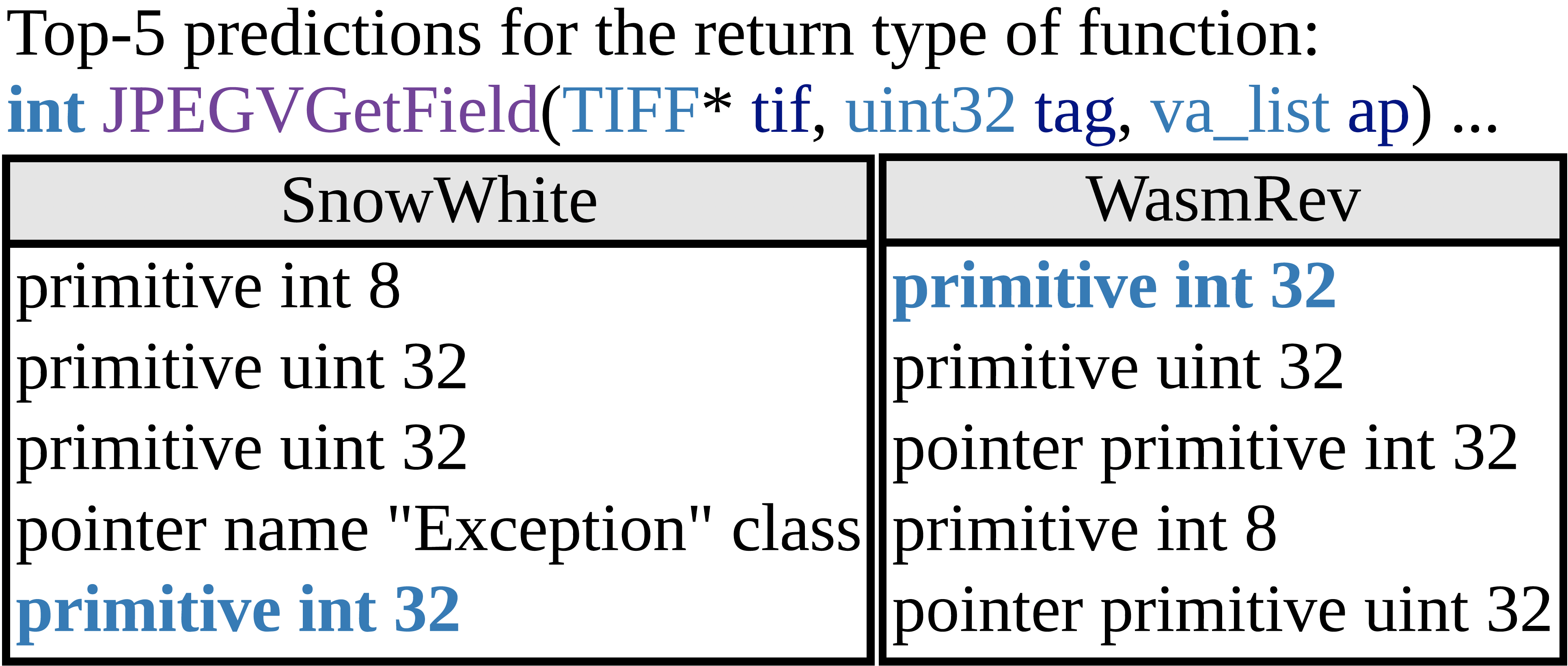}
  \captionof{figure}{A case study of TR.}
  \label{fig:tr_casestudy}
\vspace{-10pt}
\end{figure}

\noindent{\textbf{$\bullet$ Case Studies.}}
We show two case studies of TR. The first involves simple
string manipulation, as shown in Figure~\ref{fig:motivationexample}. Here \sys{} accurately identifies the parameter type as \textsf{pointer primitive char}. The second case, illustrated in Figure~\ref{fig:tr_casestudy}, showcases the top five predictions for the return type of the \textsf{JPEGVGetField} function within the \textsf{libtiff} library. In the results from SnowWhite, the correct prediction ranks fifth. In contrast, \sys{} precisely predicts the return type as \textsf{primitive int 32} in its top-1 prediction, and the other predictions within the top-5 are also close to the ground truth. 

\begin{tcolorbox}[
  enhanced,
  colback=blue!4!white,
  boxrule=0.8 pt, 
  boxsep=0pt, 
  left=2pt, 
right=2pt, 
  top=2pt, 
  bottom=2pt, 
]
\textbf{Result-2:} \sys{} is able to accurately learn the relationship of
WebAssembly and high-level programming language, and forecast precise parameter and return types from WebAssembly with high top-1/top-5 accuracy 63.7\%/88.3\% and 74.9\%/93.8\%, 
outperforming SnowWhite by 43.1\%/17.4\% and 29.8\%/16.5\% correspondingly.
\end{tcolorbox}

\subsection{\textbf{WebAssembly Summarization (RQ3)}}
To explore whether \sys{} can precisely understand the relationship of WebAssembly and natural language (RQ3), we fine-tune the pre-trained \sys{} model and evaluate it on the WebAssembly summarization (WS) task as introduced in Section~\ref{sec:ws}, using the fine-tuning dataset detailed in Section~\ref{sec:ws-dataset}. WS is a sequence generation task, where the input is a WebAssembly function and output is a sequence of natural language summaries describing that function, thereby testing model's capacity to bridge the gap between complex, low-level code constructs and their high-level, linguistic descriptions. 

\noindent{\textbf{$\bullet$ Baseline.}} As there is no existing learning-based solution for WebAssembly summarization, we train a supervised learning model with the same architecture as \sys{} on the WS dataset as our baseline, referred to as \sys{}\textsubscript{SuV}(WS).

\noindent{\textbf{$\bullet$ Metrics.}}
We use the widely-used bilingual evaluation understudy (BLEU) score~\cite{lin2004orange}, and BERTScore F1 score~\cite{zhang2019bertscore} to compare the generated sentence to reference sentence. 
Different from the normal F1 score, BERTScore utilizes the contextual embeddings from the BERT model, and compares the semantic similarity of texts by calculating the cosine similarity between the embeddings of the generated and reference texts. 
The higher BLEU score/BERTScore F1 score corresponds to a generated sentence that is more similar to the reference. 

\noindent{\textbf{$\bullet$ Results.}}
We show the results in Table~\ref{tab:FPI}. \sys{} outperforms the supervised learning baseline by 3.9\% and 4.8\% in BLEU-4 score and BERTScore F1 score, revealing \sys{}'s superior ability on understanding WebAssembly and describing it in human-readable summaries. 

\begin{figure}[h]
\centering
  \includegraphics[width=0.9\linewidth]{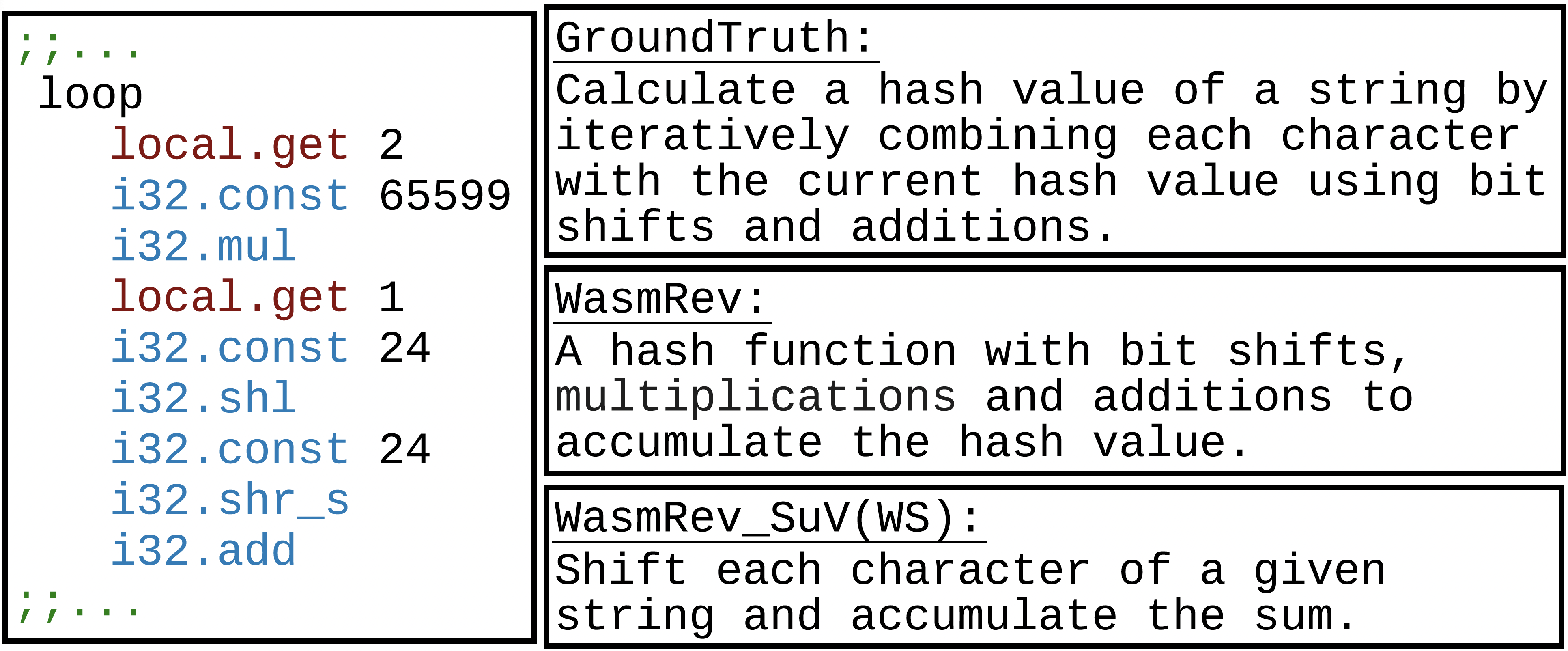}
  \captionof{figure}{A case study of WS.}
  \label{fig:casestudy}
\vspace{-10pt}
\end{figure}

\noindent{\textbf{$\bullet$ Case Studies.}} We present two case studies of WS. One is a simple string manipulation shown in Figure~\ref{fig:motivationexample}, where the WS result of \sys{} is most the same as the ground-truth documentation and presents the correct semantics. In this example, \sys{}\textsubscript{SuV}(WS) generates almost the same summarization: ``Convert the lowercase letters to uppercase.'', which is also semantically correct. Another case is shown in Figure~\ref{fig:casestudy} with partial WebAssembly code. It is a more complex hash function combining bit-wise operations and arithmetic operations, with a very detailed documentation describing its functionality. \sys{} precisely predicts the function purpose -- ``a hash function'', and correctly includes operations like ``bit shifts'' and ``additions''. It incorrectly includes ``multiplications''. We guess a possible reason is the appearance of the ``i32.mul'' between parameters 1 and 2 (``local.get 1/2'') in the WebAssembly code (as shown in Figure~\ref{fig:casestudy}) makes the model predict the ``multiplications''. The supervised learning baseline \sys{}\textsubscript{SuV}(WS) also forecasts the ``shift'' and ``accumulate'' operations but with the wrong object -- it is the hash value that is bit shifted, not the characters of the string. 
In this case, \sys{} predicts more meaningful tokens and produces a semantically closer summarization to the ground-truth. 

\vspace{-10pt}

\begin{tcolorbox}[
  enhanced,
  colback=blue!4!white,
  boxrule=0.8 pt, 
  boxsep=0pt, 
  left=2pt, 
right=2pt, 
  top=2pt, 
  bottom=2pt, 
]
\textbf{Result-3:} \sys{} is able to accurately summarize the semantics of WebAssembly in natural language, with a high BLEU-4 score (20.33) and a high BERTScore F1 (0.873) compared to non-pretrained model.
\end{tcolorbox}


\subsection{Effectiveness of multi-model inputs (RQ4)} \label{sec:effectiveness}
To evaluate the effectiveness of using multi-modal inputs, we train the following variants for each task: (1)  \sys{}\textsubscript{w/oNL} uses only (PL, Wasm) as input and accordingly removes the NL samples in SSI task; (2)  \sys{}\textsubscript{w/oPL} uses only (NL, Wasm) as input and accordingly removes the PL samples in SSI task; (3)  \sys{}\textsubscript{w/oNL+PL} uses only WebAssembly as input and uses only the different compiled versions of WebAssembly code as the similar semantics in SSI task. Other settings are identical to \sys{}. The results are shown in Table~\ref{tab:FPI} for FPI, TR and WS tasks.

\sys{} outperforms all the variants for three tasks, indicating the effectiveness of each modality of input representation. We observe removing various modalities affect differently for different tasks. 
(i) For FPI, removing NL, PL both leads to performance degradation while removing NL+PL incurs the biggest accuracy drop (-0.63\%), indicating incorporating NL and PL are both important and helpful for \sys{} to learn WebAssembly semantics. 
(ii) For TR, removing PL leads to a greater performance degradation (-0.9\%/-0.8\%) than removing only NL (-0.5\%/-0.4\%), for parameter/return TR tasks, indicating learning PL and WebAssembly concurrently helps modeling PL-WebAssembly relationship, contributing to a promising result of Wasm-to-PL task. 
(iii) For WS, removing NL leads to a greater performance degradation (-0.76) than removing only PL (-0.45), revealing that co-learning NL and Wasm helps the model grasp high-level semantics and NL-Wasm relationship. 


\begin{tcolorbox}[
  enhanced,
  colback=blue!4!white,
  boxrule=0.8 pt, 
  boxsep=0pt, 
  left=2pt, 
right=2pt, 
  top=2pt, 
  bottom=2pt, 
]
\textbf{Result-4:} Our results empirically demonstrate that incorporating various modalities of code into \sys{} contributes to improved model performance for various reverse engineering tasks.
\end{tcolorbox}

\subsection{Effectiveness of pre-training tasks (RQ5)} 
Pre-training is a critical process in which \sys{} learns multi-modal representations and cross-modal  relationships. To evaluate the effectiveness of our designed pre-training tasks, we train three kinds of variants, each of which removes one of the pre-training task T, denoted as \sys{}\textsubscript{w/oT}, T $\in$ \{M3MLM, SSI, RII\}. 

The results are shown Table~\ref{tab:FPI}, where \sys{} outperforms all the variants, showing the effectiveness of each of our design pre-training tasks. Specifically, 
(i) For FPI, removing any of the pre-training task incur performance degradation, while removing RII has a relative greater affect to FPI task. 
(ii) For TR, removing M3MLM leads to the biggest accuracy drop. The M3MLM task, which involves predicting masked source code based on its context and the semantically equivalent WebAssembly, effectively enables the model to learn the PL-Wasm relationship.
(iii) For WS, similarly, M3MLM brings the greatest impact since the model learns crucial NL-Wasm relationship through M3MLM, which is valuable for WS.


\begin{tcolorbox}[
  enhanced,
  colback=blue!4!white,
  boxrule=0.8 pt, 
  boxsep=0pt, 
  left=2pt, 
right=2pt, 
  top=2pt, 
  bottom=2pt, 
]
\textbf{Result-5:} We empirically show that our designed pre-training tasks yield promising results for various WebAssembly tasks.
\end{tcolorbox}

\subsection{\textbf{Data Efficiency (RQ6)}}
To evaluate the data efficiency of fine-tuning, i.e., the ability to achieve high model performance with a smaller amount of labeled data, we train \sys{} variants using only $\mathbf{\frac{1}{10}}$ of labeled data for \sys{} fine-tuning. We show the results in Table~\ref{tab:FPI}. For FPI and TR, although \sys{}\textsubscript{(1/10)} exhibits a minor reduction in performance compared to \sys{}, it still surpasses  baselines that utilize the full dataset for training. For WS, \sys{}\textsubscript{(1/10)} attains a BLEU-4 score and a BERTScore F1 that are both nearly on par with, yet marginally below that of \sys{}\textsubscript{SuV}(WS). Notably, \sys{}\textsubscript{(1/10)} yields a higher BLEU-4 and a higher BERTScore F1 than \sys{}\textsubscript{SuV(1/10)}(WS) that uses the same $\mathbf{\frac{1}{10}}$ of labeled data for supervised training. 
These findings underscore \sys{}'s capacity to perform efficiently in scenarios where obtaining labeled data is either challenging or resource-intensive, thereby addressing the limitation \textbf{L1} outlined in Section~\ref{sec:limitOfSOTA} and highlighting the practical advantages of \sys{} in data-constrained environments.

\begin{tcolorbox}[
  enhanced,
  colback=blue!4!white,
  boxrule=0.8 pt, 
  boxsep=0pt, 
  left=2pt, 
right=2pt, 
  top=2pt, 
  bottom=2pt, 
]
\textbf{Result-6:} Even with significantly less (\textbf{1/10}) labeled data for fine-tuning, \sys{} still achieves high model performance and outperforms the SOTA WASPur and SnowWhite.
\end{tcolorbox}

\subsection{Time Efficiency}
We discuss the time efficiency across three stages of \sys{}. 
(1) The model pre-training takes $\sim 1.5$ days on GPUs. Note that the training efficiency can be further improved by using more advanced machines such as Nvidia A100 or H100 GPUs~\cite{nvidia,nvidia-h100}, and training techniques such as mixed precision training~\cite{mix}, with a reported 1.2 hours for BERT model training~\cite{nvidia_bert}. 
The time cost of pre-training is a \textit{one-time} overhead, which is more efficient than conventional WebAssembly analysis tool development that can often take expert engineers a few weeks or months. It is more efficient than supervised training, which requires more labeled data collection effort and fully training of each model for individual tasks, resulting in a total time overhead scaling linearly with the number of tasks. 
In this sense, \sys{} improves WebAssembly tool development efficiency 
compared to conventional methods and supervised learning methods. 
(2) In the fine-tuning stage, it takes 0.5 / 9 / 1.5 hours to fine-tune \sys{} on FPI / TR / WS task. This overhead depends greatly on the size of fine-tuning dataset. 
(3) The model inference takes on average 350ms per input binary in our testing, depending on the input sequence length. Such fast inference further enhances the practicality of these reverse engineering tools developed based on \sys{}.
\vspace{-10pt}
\section{Related Work}\label{sec:related}
\noindent\textbf{Conventional WebAssembly analysis.}
Since the introduction of the WebAssembly standard~\cite{WebAssembly_website,haas2017bringing}, several studies have been conducted for WebAssembly security~\cite{konoth2018minesweeper,lehmann2020everything,musch2019new}, performance~\cite{jangda2019not}, and use in practice~\cite{hilbig2021empirical}.  
Techniques to analyze and inspect WebAssembly code include 
static analysis~\cite{brito2022wasmati}, dynamic analysis~\cite{lehmann2019wasabi} and taint analyses~\cite{fu2018taintassembly,szanto2018taint}. 
WABT~\cite{wabt} supports converting WebAssembly code to C source code and decompiling WebAssembly code into C-like syntax. 
However, none of them targets high-level semantics recovery, e.g., types, function purpose and code documentation. 

\noindent\textbf{ML for WebAssembly reverse engineering.}
To the best of our knowledge, there are very few previous works~\cite{lehmann2022finding,romano2023automated} targeting ML for WebAssembly reverse engineering. SnowWhite~\cite{lehmann2022finding} learns WebAssembly sequences and focuses on type recovery. WASPur~\cite{romano2023automated} extracts control-flow graphs of WebAssembly and targets function purpose classification. These ML methods focus on specific applications while \sys{} learns a generic WebAssembly representation for reverse engineering tasks. Recent SOTA large language models (LLMs) such as ChatGPT~\cite{ChatGPT} are not designed for WebAssembly analysis and commercial LLMs are not always viable options for WebAssembly analysis due to potential privacy concerns.
To the best of our knowledge, \sys{} is the first pre-trained language model for generalized WebAssembly reverse engineering tasks.

\noindent\textbf{Neural models of code.}
DL techniques have been extensively explored for code-related tasks, such as code summarization~\cite{leclair2019neural,wan2018improving}, vulnerability detection~\cite{li2021sysevr,szanto2018taint,fu2018taintassembly} and binary similarity comparison~\cite{diffliu2018,tian2021bindeep,yang2021asteria}.  
One avenue is to learn code representation, i.e., learn to estimate the statistical distributional properties over large and representative corpora. Earlier approaches apply 
word2vec~\cite{mikolov2013efficient} to obtain code representations~\cite{bojanowski2017enriching,chen2019literature}. More recent works exploit representation learning on SQL query~\cite{jain2018query2vec}, source-level programming languages~\cite{dam2016deep,gu2016deep}, binary code~\cite{wang2022jtrans,pei2021stateformer}, and bi-modal (NL-PL)~\cite{feng2020codebert, guo2020graphcodebert}. 
Our exploration resonates with recent studies on the principle of neural software analysis~\cite{NeuralSE} and code representation learning~\cite{niu2022deep}, while proposing the first multi-modal representation learning of NL-PL-Wasm for generalized WebAssembly tasks.

\vspace{-4pt}
\section{Conclusion}\label{sec:conclusion}
This paper proposes \sys{}, the first multi-modal pre-trained language model, which learns a generic representation of source code, code documentation, and WebAssembly code. It can be applied to diverse WebAssembly reverse engineering tasks. 
Incorporating multi-modal inputs and tailored pre-training tasks, \sys{} can learn from cross-modal relationships and gain a 
holistic understanding of WebAssembly. 
\sys{} generates accurate results for type recovery, function purpose identification, and WebAssembly summarization. 
It outperforms supervised learning methods, providing developers with high-level semantics for inspecting WebAssembly modules and assisting WebAssembly comprehension. 
We leave exploring \sys{} for 
other reverse engineering tasks such as function name prediction and precise decompilation as future work.

\vspace{-4pt}
\section{Acknowledgment}
The authors thank the anonymous reviewers for their valuable feedback and comments. This paper is supported in part by NSF grants 1829524, 1817077, 2011212, and the PRISM center in JUMP 2.0, an SRC program sponsored by DARPA.

\balance
\bibliographystyle{ACM-Reference-Format}
\bibliography{sample-base}

\end{document}